\documentstyle[12pt,a4]{article}

\begin{document}

\newcommand{\nc}[2]{\newcommand{#1}{#2}}
\newcommand{\ncx}[3]{\newcommand{#1}[#2]{#3}}
\ncx{\pr}{1}{#1^{\prime}}
\nc{\nl}{\newline}
\nc{\np}{\newpage}
\nc{\nit}{\noindent}
\nc{\be}{\begin{equation}}
\nc{\ee}{\end{equation}}
\nc{\ba}{\begin{array}}
\nc{\ea}{\end{array}}
\nc{\bea}{\begin{eqnarray}}
\nc{\eea}{\end{eqnarray}}
\nc{\nb}{\nonumber}
\nc{\dsp}{\displaystyle}
\nc{\bit}{\bibitem}
\nc{\ct}{\cite}
\ncx{\dd}{2}{\frac{\partial #1}{\partial #2}}
\nc{\pl}{\partial}
\nc{\dg}{\dagger}
\nc{\cH}{{\cal H}}
\nc{\cL}{{\cal L}}
\nc{\cD}{{\cal D}}
\nc{\cF}{{\cal F}}
\nc{\cG}{{\cal G}}
\nc{\cJ}{{\cal J}}
\nc{\cQ}{{\cal Q}}
\nc{\tB}{\tilde{B}}
\nc{\tD}{\tilde{D}}
\nc{\tH}{\tilde{H}}
\nc{\tR}{\tilde{R}}
\nc{\tZ}{\tilde{Z}}
\nc{\tg}{\tilde{g}}
\nc{\tog}{\tilde{\og}}
\nc{\tGam}{\tilde{\Gam}}
\nc{\tPi}{\tilde{\Pi}}
\nc{\tcD}{\tilde{\cD}}
\nc{\tcQ}{\tilde{\cQ}}
\nc{\ttau}{\tilde{\tau}}
\nc{\ag}{\alpha}
\nc{\bg}{\beta}
\nc{\gam}{\gamma}
\nc{\Gam}{\Gamma}
\nc{\bgm}{\bar{\gam}}
\nc{\del}{\delta}
\nc{\Del}{\Delta}
\nc{\eps}{\epsilon}
\nc{\ve}{\varepsilon}
\nc{\zg}{\zeta}
\nc{\th}{\theta}
\nc{\vt}{\vartheta}
\nc{\Th}{\Theta}
\nc{\kg}{\kappa}
\nc{\lb}{\lambda}
\nc{\Lb}{\Lambda}
\nc{\ps}{\psi}
\nc{\Ps}{\Psi}
\nc{\sg}{\sigma}
\nc{\spr}{\pr{\sg}}
\nc{\Sg}{\Sigma}
\nc{\rg}{\rho}
\nc{\fg}{\phi}
\nc{\Fg}{\Phi}
\nc{\vf}{\varphi}
\nc{\og}{\omega}
\nc{\Og}{\Omega}
\nc{\dL}{\del_{\Lambda}}
\nc{\Kq}{\mbox{$K(\vec{q},t|\pr{\vec{q}\,},\pr{t})$ }}
\nc{\Kp}{\mbox{$K(\vec{q},t|\pr{\vec{p}\,},\pr{t})$ }}
\nc{\vq}{\mbox{$\vec{q}$}}
\nc{\qp}{\mbox{$\pr{\vec{q}\,}$}}
\nc{\vp}{\mbox{$\vec{p}$}}
\nc{\va}{\mbox{$\vec{a}$}}
\nc{\vb}{\mbox{$\vec{b}$}}
\nc{\Ztwo}{\mbox{\sf Z}_{2}}
\nc{\Tr}{\mbox{Tr}}
\nc{\lh}{\left(}
\nc{\rh}{\right)}
\nc{\ld}{\left.}
\nc{\rd}{\right.}
\nc{\nil}{\emptyset}
\nc{\bor}{\overline}
\nc{\ha}{\hat{a}}
\nc{\da}{\hat{a}^{\dg}}
\nc{\hb}{\hat{b}}
\nc{\db}{\hat{b}^{\dg}}
\nc{\hc}{\hat{c}}
\nc{\he}{\hat{e}}
\nc{\hp}{\hat{p}}
\nc{\hx}{\hat{x}}
\nc{\hH}{\hat{H}}
\nc{\hI}{\hat{I}}
\nc{\hK}{\hat{K}}
\nc{\hM}{\hat{M}}
\nc{\hN}{\hat{N}}
\nc{\hQ}{\hat{Q}}
\nc{\hag}{\hat{\ag}}
\nc{\hbg}{\hat{\bg}}
\nc{\hgm}{\hat{\gam}}
\nc{\hch}{\hat{\chi}}
\nc{\hps}{\hat{\psi}}
\nc{\hDel}{\hat{\Del}}
\nc{\hOg}{\hat{\Og}}
\nc{\pz}{z^{\prime}}
\nc{\pZ}{Z^{\prime}}
\ncx{\abs}{1}{\left| #1 \right|}
\nc{\vs}{\vspace{2ex}}

\pagestyle{empty}

\begin{flushright}
\begin{tabular}{c}
NIKHEF/95-055 \\
hep-th/9510021 \\
\end{tabular}
\end{flushright}

\begin{center}
{\LARGE {\bf $D = 1$ Supergravity and Spinning Particles$^*$  }} \\

\vspace{5ex}

{\large J.W.\ van Holten} \\
        NIKHEF/FOM, P.O.\ Box 41882 \\
        1009 DB Amsterdam NL \\

\today \\

\vspace{12ex}

{\bf Abstract} \\
\end{center}

\nit
{\small
In this paper I review the multiplet calculus of $N = 1$, $D = 1$ local
supersymmetry with applications to the construction of models for spinning
particles in background fields, and models with space-time supersymmetry.
New features include a non-linear realization of the local supersymmetry
algebra and the coupling to anti-symmetric tensor fields of both odd and
even rank. The non-linear realization allows the construction of a $D = 1$
cosmological-constant term, which provides a mass term in the equations of
motion.
}
\vfill

\nit
{\footnotesize $^*$ Dedicated to Jurek Lukierski on the occasion of his 60th
anniversary.}

\np

\pagestyle{plain}
\pagenumbering{arabic}

\section{Worldline supersymmetry}{\label{S.1}}

Supersymmetry, as a symmetry between bosons and fermions, was discovered
almost 25 years ago \ct{golf}-\ct{wesz}. Apart from mathematical
elegance, supersymmetry has the quality of improving the short-distance
behaviour of quantum theories and has therefore been proposed as an
ingredient of many models of physical phenomena, most often but not
exclusively in the domain of particle physics and quantum gravity. Many of
these applications are presently speculative, but it was realized already
early that supersymmetric extensions of relativistic particle mechanics
describe ordinary Dirac fermions \ct{BerMar}-\ct{BCL}. Supersymmetric
theories of this type are known as spinning particle models. They are
useful providing low-energy descriptions for fermions in external fields
\ct{BCL}-\ct{jw10} and path-integral expressions for perturbative amplitudes
in quantum field theories \ct{shvarts}-\ct{pvn}. They are also useful in
studying aspects of higher-dimensional supersymmetric field theories and
superstring models \ct{GSW}.

Like in string theory, in the discussion of supersymmetric point-particle
models one has to distinguish between worldline supersymmetry, where the
supersymmetry concerns transformations of the worldline parameters of the
physical variables (proper time), and space-time supersymmetry which
refers to supersymmetry in the target space of the physical variables.
Both types of supersymmetry are encountered in the literature. In fact,
in \ct{lukjw} a model was constructed possessing both types of supersymmetry
simultaneously.

In the present paper I review the construction of pseudo-classical spinning
particle models with worldline supersymmetry. To this end I present the
multiplet calculus for local worldline supersymmetry ($N=1$ supergravity in
one dimension) and construct general lagrangians for $D=1$ supersymmetric
linear and non-linear $\sg$-models with potentials; many elements of this
formalism were developed in \ct{RR,jacq}. The coupling to all kinds of
background fields, including scalars, abelian and non-abelian vectors fields,
gravity and anti-symmetric tensors is discussed in some detail. I finish with
the construction of a model which exhibits space-time supersymmetry as well
\ct{lukjw}.

\section{$D = 1$ supermultiplets}{\label{S.8}}

Supergravity models in 0 + 1 space-time dimensions describe spinning
particles. Indeed, the local supersymmetry and reparametrization invariance
generate first-class constraints which after quantization can be identified
with the Dirac and Klein-Gordon equations. Hence the quantum states of the
model are spinorial wave functions for a fermion in a $d$-dimensional target
space-time.

The models I construct below are general $N = 1$ supergravity actions in
$D = 1$ with at most 2 proper-time derivatives, and the correponding quantum
theories. Higher-$N$ models have been considered for example in
\ct{jw11,sorok}. Extended target-space supersymmetry has been studied in
\ct{azc,fryd}.

In one dimension supersymmetry is realized off-shell\footnote{The term
{\em off shell} implies that the supersymmetry algebra is realized without
using dynamical constraints like the equations of motion.} by a number of
different sets of variables, the supermultiplets (superfields):

\begin{enumerate}

\item{ The gauge multiplet $(e, \chi)$ consisting of the einbein $e$ and
its superpartner $\chi$, the gravitino; under infinitesimal local worldline
super-reparametrizations, generated by the parameter-valued operator
$\del(\xi,\ve)$ where $\xi(\tau)$ is the commuting parameter of translations
and $\ve(\tau)$ the anti-commuting parameter of supersymmetry, the multiplet
transforms as

\be
\del e\, =\, \frac{d(\xi e)}{d\tau} - 2 i \ve \chi, \hspace{3em}
\del \chi\, =\, \frac{d(\xi \chi)}{d\tau} + \frac{d\ve}{d\tau}.
\label{8.1}
\ee
}

\item{ Scalar multiplets $(x, \psi)$, used to describe the position
and spin co-ordinates of particles. The transformation rules are

\be
\del x\, =\, \xi \frac{dx}{d\tau} - i \ve \psi, \hspace{3em}
\del \psi\, =\, \xi \frac{d\psi}{d\tau} + \ve \frac{1}{e} \cD_{\tau} x,
\label{8.2}
\ee

\nit
where the supercovariant derivative is constructed with the gravitino
as the connection:

\be
\cD_{\tau} x\, =\, \frac{dx}{d\tau}\, +\, i \chi \psi.
\label{8.3}
\ee
}

\item{ Fermionic multiplets $(\eta, f)$ with Grassmann-odd $\eta$ and
even $f$. The $f$-component is most often used as an auxiliary variable,
without dynamics of itself. The transformation properties under local
super-reparametrizations are

\be
\del \eta\, =\,  \xi \frac{d\eta}{d\tau} + \ve f, \hspace{3em}
\del f\, =\,  \xi \frac{df}{d\tau} - i \ve \frac{1}{e} \cD_{\tau} \eta.
\label{8.4}
\ee

\nit
The supercovariant derivative is formed using the same recipe as before:

\be
\cD_{\tau} \eta\, =\, \frac{d\eta}{d\tau}\, -\, \chi f.
\label{8.5}
\ee
}

\item{ A non-linear multiplet consisting of a single fermionic component
$\sg$ with the transformation rules

\be
\del \sg\, =\, \xi \frac{d\sg}{d\tau} + \ve -
               i \ve \frac{1}{e} \sg \cD_{\tau} \sg,
\label{8.6}
\ee

\nit
with the supercovariant derivative

\be
\cD_{\tau} \sg\, =\, \frac{d\sg}{d\tau} - \chi +
                     \frac{i}{e} \chi \sg \frac{d\sg}{d\tau}.
\label{8.7}
\ee
}

\end{enumerate}

\nit
On any component of any multiplet the commutator of two infinitesimal
variations with parameters $(\xi_{1,2},\ve_{1,2})$ results in an
infinitesimal transformation with parameters $(\xi_3,\ve_3)$ given by

\be
\ba{c}
\left[ \del(\xi_2,\ve_2), \del(\xi_1,\ve_1) \right] = \del(\xi_3,\ve_3), \\
  \\
\dsp{
\xi_3 = \xi_1 \frac{d\xi_2}{d\tau} - \xi_2 \frac{d\xi_1}{d\tau} -
        \frac{2i}{e} \ve_1 \ve_2, } \\
  \\
\dsp{
\ve_3 = \xi_1 \frac{d\ve_2}{d\tau} - \xi_2 \frac{d\ve_1}{d\tau} +
        \frac{2i}{e} \ve_1 \ve_2 \chi. }
\ea
\label{8.5.1}
\ee

\nit
For the non-linear representation $\sg$ the proof requires use of the
supersymmetry variation

\be
\del(\ve) \lh \frac{1}{e} \cD_{\tau} \sg \rh\, =\, - i \ve
  \frac{1}{e} \cD_{\tau} \lh \frac{1}{e} \sg \cD_{\tau} \sg \rh\, =\,
  - i \ve \frac{1}{e} \sg \cD_{\tau} \lh \frac{1}{e} \cD_{\tau} \sg \rh.
\label{8.5.2}
\ee

\nit
Then a simple result is obtained:

\be
\del(\ve) \lh \frac{1}{e} \sg \cD_{\tau} \sg \rh\, =\,
     \ve \frac{1}{e} \cD_{\tau} \sg.
\label{8.5.3}
\ee

\nit
It follows that each of these multiplets is a representation of the same
local supersymmetry algebra, and this algebra closes off-shell. However,
the parameters of the resulting transformation depend on the components
of the gauge multiplet $(e, \chi)$, indicating that the algebra of
infinitesimal transformations is a {\em soft} commutator algebra, rather
than an ordinary super Lie-algebra.

Among the representations discussed, the gauge multiplet and the non-linear
multiplet have manifestly non-linear transformation rules. The variations of
the other two multiplets are linear in the components of these multiplets.
For this reason the scalar and fermionic multiplets are called {\em linear}
representations of local supersymmetry, although some of the coefficients
depend on the gauge variables $(e, \chi)$.

\section{Multiplet calculus}{\label{S.3}}

The linear representations (scalar and fermionic) satisfy some simple
addition and multiplication rules; this tensor calculus has been developed
in \ct{jacq}. The rules can also be formulated in terms of $D = 1$
superfields \ct{RR}. As concerns addition, any linear multiplets of the
same type can be added component by component with arbitrary real or
complex coefficients. The linearity of the transformation rules then
guarantees the sum to be a multiplet of the same type.

The multiplication rules are also simple. There are 3 different product
formula's:

\begin{enumerate}

\item{The product of two scalar multiplets $\Sg = (x, \psi)$,
$\Sg^{\prime} = (x^{\prime}, \psi^{\prime})$ is a scalar multiplet

\be
\Sg \times \Sg^{\prime} = \Sg^{\prime\prime} = \lh x x^{\prime},
    x \psi^{\prime} + x^{\prime} \psi \rh.
\label{3.1}
\ee

\nit
This rule can be extended to arbitrary powers of scalar multiplets, for
example:

\be
\Sg^n = \lh x^n, n x^{n-1} \psi \rh.
\label{3.1.1}
\ee

\nit
In this way one can define functions of scalar multiplets by power series
expansions.
}

\item{The product of a scalar multiplet $\Sg = (x, \psi)$ and a fermionic
multiplet $\Fg = (\eta, f)$ is a fermionic multiplet

\be
\Sg \times \Fg = \Fg^{\prime} = \lh x \eta, x f - i \psi \eta \rh.
\label{3.2}
\ee
}

\item{The product of two fermionic multiplets is a scalar multiplet:

\be
-i \Fg \times \Fg^{\prime} = \Sg^{\prime} =
           \lh -i \eta \eta^{\prime}, f \eta^{\prime} - f^{\prime} \eta \rh.
\label{3.3}
\ee
}

\end{enumerate}

\nit
Next I introduce the operation of derivation of scalar and fermionic
multiplets; the super-derivative on linear multiplets is a Grassmann-odd
linear operator turning a scalar multiplet into a fermionic multiplet,
and vice-versa, with the following components:

\be
\ba{l}
\cD \Sg = \Fg^{\prime} = \lh \psi, \frac{1}{e} \cD_{\tau} x \rh; \\
 \\
\cD \Fg = \Sg^{\prime} = \lh f, \frac{1}{e} \cD_{\tau} \eta \rh.
\ea
\label{3.4}
\ee

\nit
On product multiplets this super-derivative satisfies the Leibniz rule,
with in particular the result

\be
\cD \Sg^n = n \cD \Sg \times \Sg^{n-1}.
\label{3.5}
\ee

\nit
The super-derivative satisfies an operator algebra similar to the
supersymmetry algebra:

\be
\cD^2 = \frac{1}{e}\, \cD_{\tau},
\label{3.6}
\ee

\nit
where $\cD_{\tau}$ is the supercovariant proper-time derivative on components,
encountered before in eqs.(\ref{8.3}) and (\ref{8.5}).

\section{Invariant actions}{\label{S.4}}

{\em 1.} Invariant actions can be constructed for the linear as well as
the non-linear multiplets. As there exists no intrinsic curvature in $D = 1$,
there is no invariant action for the gauge multiplet involving the einbein,
but there is a very simple action for the gravitino, namely

\be
S_{\Lb}\, =\, \int d\tau\, i \Lb \chi,
\label{4.0}
\ee

\nit
where $\Lb$ is a constant. The equation of motion for this action by
itself is not consistent (it requires $\Lb$ to vanish); but this is
changed if one adds other terms to the action, like the ones discussed
below. Also, one can replace $\chi$ in the action by $d\chi/d\tau$, but
then the action becomes a total derivative.

A cosmological-constant like action can be constructed with the help
of the non-linear multiplet; it reads

\be
S_{nl}(\sg)\, =\,
     \int d\tau\, \lh e - i \chi \sg - i \sg \cD_{\tau} \sg \rh.
\label{4.1}
\ee

\nit
This is also the kinetic action for the fermionic $\sg$ variable, which
in view of the anti-commuting nature of $\sg$ can be only linear in
proper-time derivatives. Note, that the non-linear nature of the
multiplet allows one to rescale the variable $\sg$ and thereby change
the relative co-efficients between the various terms in the action. A
rescaling of $\sg$ by a factor $1/c$ gives the action

\be
S_{nl}(\sg;c)\, =\, \int d\tau\, \lh
  e - \frac{2i}{c}\, \chi \sg - \frac{i}{c^2}\, \sg \dot{\sg} \rh,
\label{4.2}
\ee

\nit
where I have introduced the dot notation for ordinary proper-time derivatives.
Of course, the rescaling also changes the non-linear transformation rule
for $\sg$ under supersymmetry to

\be
\del_{c} \sg\, =\, \xi \frac{d\sg}{d\tau} + c\, \ve -
               \frac{i \ve}{c e} \sg \cD_{\tau} \sg.
\label{4.3}
\ee

\nit
Combining the actions $S_{\Lb}$ and $S_{nl}$ in such a way as to get standard
normalization of the fermion kinetic term for $\sg$ leads to the
action (with the dimension of $\hbar$)

\be
\ba{lll}
S_{grav} & = & \dsp{ mc S_{\Lb} - \frac{mc^2}{2}\, S_{nl} } \\
 & & \\
 & = & \dsp{ m \int d\tau\, \lh i c \Lb \chi - \frac{c^2}{2}\, e + i c \chi \sg
             + \frac{i}{2}\, \sg \dot{\sg} \rh, }
\ea
\label{4.3.0}
\ee

\nit
where $m$ is a parameter with the dimension of mass and $c$ has the dimension
of a velocity. The Euler-Lagrange equations for the fermions $\chi$ and $\sg$
then give

\be
\sg = \Lb, \hspace{3em} \chi = \frac{1}{c}\, \frac{d\sg}{d\tau} = 0.
\label{4.3.1}
\ee

\nit
Thus the constant $\Lb$ is like a vacuum expectation value of the fermionic
variable $\sg$. However, the variation (\ref{4.3}) of $\sg$ is such that for
non-zero $c$ it can be gauged away completely by a supersymmetry
transformation. Therefore it does not represent a true physical degree of
freedom. Of course, this seems to contradict the equation of motion
(\ref{4.3.1}), but we observe that also the equation for the einbein is
inconsistent, requiring the constant $c$ to vanish. Again, these problems
are solved by adding further terms to the action. In applications $c$ usually
represents the velocity of light, which can conveniently be taken as unity
$(c = 1)$.
 \vs

\nit
{\em 2.} Next we turn to a formula for the construction of invariant
actions for linear multiplets. Given a fermionic multiplet $\Fg = (\eta, f)$,
an action invariant under local supersymmetry transformations is

\be
S_{lin} = \int d\tau\, \lh e f - i \chi \eta \rh.
\label{3.7}
\ee

\nit
Note, that in eqs.(\ref{4.1}) and (\ref{3.7}) the integrand itself is
not invariant, but transforms into a total proper-time derivative:

\be
\del S_{lin} = \int d\tau \frac{d}{d\tau} \lh -i \ve \eta \rh,
\label{3.8}
\ee

\nit
and the same for $\del S_{nl}$ with $\eta \rightarrow \sg$. Eq.(\ref{3.8})
shows, that $\del S$ vanishes for variations which are zero on the endpoints.

Eq.(\ref{3.7}) can be applied to the construction of actions for scalar
multiplets if one applies an odd number of super-derivatives so as to
obtain a composite fermionic multiplet, sometimes called the (fermionic)
prepotential. A simple example of this construction is the free kinetic
action for a scalar multiplet, constructed from the composite fermionic
multiplet

\be
\Fg_{kin} = \frac{1}{2}\, \cD^2 \Sg \times \cD \Sg,
\label{3.9}
\ee

\nit
Inserting the components of this multiplet into the action formula
(\ref{3.7}) gives

\be
S_{kin} = \int d\tau\, \lh \frac{1}{2e}\, \dot{x}^2
        + \frac{i}{2}\, \psi \dot{\psi} + \frac{i}{e}\, \chi \psi \dot{x} \rh.
\label{3.10}
\ee

\nit
If one extends this formula to $d$ free multiplets $\Sg^{\mu} =
(x^{\mu}, \psi^{\mu})$, $\mu = 1,...,d$, then using the appropriate
minkowski metric it becomes the action for a free spinning particle
in $d$-dimensional space-time.

This is a special case of the most general action involving only
scalar multiplets and quadratic in proper-time derivatives of the
bosonic co-ordinates $\dot{x}^{\mu}$: the $D = 1$ (non-)linear $\sg$-model
in a $d$-dimensional target space, constructed from a fermionic multiplet

\be
\Fg \left[ g \right]\, =\, \frac{1}{2}
           g_{\mu\nu}(\Sg) \times \cD^2 \Sg^{\mu} \times \cD \Sg^{\nu}.
\label{4.4}
\ee

\nit
Here $g_{\mu\nu}(\Sg)$ is a symmetric tensor in the space of the scalar
multiplets, which can be interpreted as a metric on the target manifold.
Using $\Fg[g]$ in the action formula (\ref{3.7}) gives the component
expression

\be
S_{kin}\left[ g \right]\, =\, \int d\tau\, \lh \frac{1}{2e} g_{\mu\nu}(x)
  \dot{x}^{\mu} \dot{x}^{\nu} + \frac{i}{2} g_{\mu\nu}(x) \psi^{\mu}
  D\psi^{\nu} + \frac{i}{e} g_{\mu\nu}(x) \chi \psi^{\mu}
  \dot{x}^{\nu} \rh,
\label{4.5}
\ee

\nit
where $D$ denotes a target-space covariant derivative

\be
D\psi^{\mu}\, =\, \dot{\psi}^{\mu} +
                  \dot{x}^{\lb} \Gam_{\lb\nu}^{\:\:\:\:\:\mu} \psi^{\nu},
\label{4.6}
\ee

\nit
with $\Gam_{\kg\nu}^{\:\:\:\:\:\mu}(x)$ the Riemann-Christoffel connection.
The action $S_{kin}[g]$ is manifestly covariant in the target space. The
symmetries of this action have been investigated in detail in \ct{rjw,grjw},
with applications to special target manifolds like Schwarzschild space-time
and Taub-NUT in \ct{grjw,rjw2,jw10}.

The simplest action for scalar multiplets involves only one super-derivative.
It starts from the general fermionic multiplet (super 1-form)

\be
\Fg \left[ A \right]\, =\, A_{\mu}(\Sg) \times \cD \Sg^{\mu},
\label{4.7}
\ee

\nit
with $A_{\mu}(\Sg)$ an abelian vector field on the target space. Inserting
the components of $\Fg[A]$ in the action formula (\ref{3.7}) gives the result

\be
S_{vec}\left[ A \right]\, =\, \int d\tau\, \lh A_{\mu}(x) \dot{x}^{\mu}
   - \frac{ie}{2}\, F_{\mu\nu}(x) \psi^{\mu}\psi^{\nu} \rh,
\label{4.8}
\ee

\nit
with $F_{\mu\nu}$ the field strength tensor of the abelian vector field
$A_{\mu}(x)$.

A similar construction can be carried out using arbitrary odd super $p$-forms
\ct{RR}. For example $p = 3$ gives

\be
\Fg \left[ H \right]\, =\, \frac{i}{3!}\, H_{\mu\nu\lb}(\Sg) \times
     \cD \Sg^{\mu} \times \cD \Sg^{\nu} \times   \cD \Sg^{\lb}.
\label{4.8.1}
\ee

\nit
This gives an action involving the anti-symmetric 3-tensor $H_{\mu\nu\lb}(x)$:

\be
S_{odd}\left[ H \right]\, =\, \frac{i}{2} H_{\mu\nu\lb}(x)
 \ps^{\mu}\ps^{\nu} \lh \dot{x}^{\lb} + \frac{2i}{3} \chi \ps^{\lb} \rh +
 \frac{e}{3!} \pl_{\kg} H_{\mu\nu\lb}(x) \ps^{\kg} \ps^{\mu} \ps^{\nu}
\ps^{\lb}.
\label{4.8.2}
\ee

\nit
The inclusion of even $p$-forms is also possible, but requires one or more
fermionic multiplets; details are given below.

In the same spirit one can find odd $p$-form extensions of the kinetic term

\be
\Fg \left[ G \right]\, =\, G_{\mu\nu_{1}...\nu_{p}}(\Sg) \times
  \cD^2 \Sg^{\mu} \times \cD \Sg^{\nu_1} ... \times \Sg^{\nu_p}.
\label{4.8.3}
\ee

\nit
For a discussion of these unconventional actions I refer to \ct{RR}. Finally,
actions with higher powers of $\cD^2$ and/or with $\cD^n$ ($n \geq 3$)
lead to higher-derivative component lagrangians. I do not consider them here.

Combining the results for scalar fields, within the restrictions we have
imposed the general action for scalar multiplets is of the form

\be
S\left[ \Sg \right]\, =\, m S_{kin}\left[ g \right] +
  q S_{vec}\left[ A \right] + \ag S_{odd} \left[ H \right] + mc
  S_{\Lb} - \frac{mc^2}{2} S_{nl}.
\label{4.9}
\ee

\nit
This action describes a spinning particle in background electro-magnetic
and gravitational fields, with the possible inclusion of torsion for $\ag
\neq 0$. The first-class constraints obtained from the equation of motion
for the einbein and gravitino are

\be
\ba{l}
\dsp{ m^2 g_{\mu\nu} \lh \dot{x}^{\mu} + i \chi \ps^{\mu} \rh
           \lh \dot{x}^{\nu} + i \chi \ps^{\nu} \rh =
           - m e^2 \lh m c^2 + iq F_{\mu\nu}(A) \ps^{\mu} \ps^{\nu}
           - \frac{\ag}{3}\, F_{\kg\mu\nu\lb}(H) \ps^{\kg}
           \ps^{\mu} \ps^{\nu} \ps^{\lb} \rh. }\\
  \\
\dsp{ m g_{\mu\nu} \dot{x}^{\mu} \ps^{\nu} - \frac{i\ag}{3} H_{\mu\nu\lb}
      \ps^{\mu} \ps^{\nu} \ps^{\lb}\, =\, mc e \lh \Lb - \sg \rh. }
\ea
\label{4.9.1}
\ee

\nit
Here $F_{\mu\nu}(A)$ and $F_{\kg\mu\nu\lb}(H)$ are the field strenghts of
the vector $A_{\mu}$ and 3-form $H_{\mu\nu\lb}$, respectively.
Eqs.(\ref{4.9.1}) are the pseudo-classical equivalents of the Klein-Gordon
and Dirac equations. Note that local supersymmetry can be used to chose a
gauge $\sg = \Lb$ in which the expressions on both sides of the second
equation vanish. If $c = 0$ (absence of $S_{nl}$ and $S_{\Lb}$) the particle
is massless. With the inclusion of $S_{nl}$  $(c \neq 0)$ the particle
acquires a non-zero mass.

If the kinetic terms are normalized in the standard way, the relative
co-efficient $q$ between the first two terms represents the electric
charge of the spinning particle, as defined by the generalized
Lorentz-force \ct{jw1a}. Then the anti-symmetric tensor $D^{\mu\nu} =
q \psi^{\mu} \psi^{\nu}$ represents the electric and magnetic dipole
moments. The terms involving the 3-form $H_{\mu\nu\lb}(x)$ combine to
form an anti-symmetric contribution to the Riemann-Christoffel
connection, representing torsion indeed.
\vs

\nit
{\em 3.} Finally we turn to the construction of actions involving
elementary fermionic multiplets. To begin with, there is the simple
action formula (\ref{3.7}) linear in the components of a single
fermionic multiplet. It involves no proper-time derivatives, and
therefore it can constribute only to potential terms. A natural and
straightforward generalization of this action involving $r$ fermionic
multiplets $\Fg^i$, $i = 1,...,r$, is constructed from the composite
fermionic prepotential

\be
\Fg \left[ U \right]\, =\, U_i(\Sg) \times \Fg^i,
\label{4.10}
\ee

\nit
with the $U_i(\Sg)$ a set of scalar-multiplet valued potentials. The
component action then is

\be
S_{pot} \left[ U \right]\, =\, \int d\tau\, \lh e U_i(x) f^i
  - i U_i(x) \chi \eta^i - i e \psi^{\mu} \pl_{\mu} U_i(x) \eta^i \rh.
\label{4.11}
\ee

\nit
As the equation of motion for $f^i$ requires all $U_i(x)$ to vanish,
this action by itself is useful only to impose constraints on the
target manifold. This conclusion is modified when additional (kinetic)
terms are added to the action.

More complicated actions obtained using the multiplet calculus with both
fermionic and scalar multiplets must have an odd total number of fermionic
multiplets and super-derivatives. Therefore the next complicated
type of action involves the product of two fermionic multiplets including
a super-derivative. The general form of the fermionic prepotential is

\be
\Fg \left[ K \right]\, =\, \frac{1}{2}\,
           K_{ij}(\Sg) \times \cD \Fg^i \times \Fg^j,
\label{4.12}
\ee

\nit
with $K_{ij}(\Sg)$ a scalar-multiplet valued symmetric matrix. The
component action for this prepotential is

\be
S_{ferm}\left[ K \right]\, =\, \int d\tau\, \lh \frac{i}{2} K_{ij}(x)
   \eta^i \dot{\eta}^j + \frac{e}{2} K_{ij}(x) f^i f^j -
   \frac{ie}{2} \psi^{\mu} \pl_{\mu} K_{ij}(x) f^i \eta^j \rh.
\label{4.13}
\ee

\nit
It contains kinetic terms for the fermionic variables $\eta^i$, but
the variables $f^i$ only appear without derivatives and are auxiliary
degrees of freedom. In combination with the potential term $S_{pot}[U]$
its elimination turns the constraints $U_i$ into a true potential,
allowing the bosonic variables to fluctuate around the solutions of
the constraints $U_i(x) = 0$.

Other actions can be constructed by replacing some of the super-derivatives
$\cD \Sg^{\mu}$ in the odd $p$-form prepotentials like $\Fg[H]$ (\ref{4.8.1})
by fermionic multiplets. I give the details for the case $p = 3$. First
consider a prepotential linear in fermionic multiplets:

\be
\Fg \left[ B \right]\, =\,  - \frac{i}{2}\, B_{i\mu\nu}(\Sg) \times \Fg^i
\times
    \cD \Sg^{\mu} \times \cD \Sg^{\nu}.
\label{4.14}
\ee

\nit
The potentials $B_{i\mu\nu}(x)$ define $r$ anti-symmetric tensors (2-forms)
on the target space of the scalars. Thus this construction and its higher-rank
generalizations allows the inclusion of even $p$-forms in the action.
Substitution in the linear-multiplet action $S_{lin}$, eq.(\ref{3.7}), gives

\be
\ba{lll}
S_{even}\left[ B \right] & = & \int d\tau\, \dsp{
  \lh - \frac{ie}{2} f^i B_{i\mu\nu}(x) \ps^{\mu} \ps^{\nu} -
  i \eta^i B_{i\mu\nu}(x) \ps^{\mu} \dot{x}^{\nu} + \frac{1}{2} \chi \eta^i
  B_{i\mu\nu}(x) \ps^{\mu} \ps^{\nu} \rd }\\  & & \\
& & \dsp{ \ld \hspace{3em} +\,
\frac{e}{2} \eta^i \pl_{\lb} B_{i\mu\nu}(x) \ps^{\lb} \ps^{\mu} \ps^{\nu} \rh.
}
\ea
\label{4.15}
\ee

\nit
Next consider the case of a quadratic expression in fermionic multiplets.
The prepotential is

\be
\Fg \left[ V \right]\, =\, \frac{i}{2}\, V_{ij\mu}(\Sg) \times \Fg^i \times
                           \Fg^j \times \cD \Sg^{\mu}.
\label{4.16}
\ee

\nit
The vector field $V_{ij\mu}(x)$, anti-symmetric in $[ij]$, takes values in a
gauge group $G \subseteq SO\lh r \rh$ for $r$ even (in the quantum theory this
is always the case \ct{jw9}). Thus this action describes the coupling to
Yang-Mills fields. After quantization the fermionic variables $\eta^i$ generate
a Clifford-algebra representation of the group $G$ embedded in $SO(r)$ on the
particle wave-functions. The explicit expression for the pseudo-classical
action is

\be
S_{YM}\left[ V \right]\, =\, \int d\tau\, \lh
  \frac{i}{2} \eta^i \eta^j V_{ij\mu}(x) \dot{x}^{\mu} +
  \frac{e}{4}\, \eta^i \eta^j F^{(0)}_{ij\mu\nu}(x) \ps^{\mu} \ps^{\nu}  +
  i e f^i V_{ij\mu}(x) \eta^j \ps^{\mu} \rh.
\label{4.17}
\ee

\nit
Here $F^{(0)}_{\mu\nu}$ represents the abelian (linear) part of the
field-strength for the vector field $V_{\mu}$. The non-abelian part
can be obtained by a proper choice of $K_{ij}$ in $S_{ferm}$ and subsequent
elimination of the auxiliary variables $f^i$ (see below).

Finally I consider the action cubic in fermionic multiplets:

\be
\Fg \left[ T \right]\, =\, \frac{i}{3!}\, T_{ijk}(\Sg) \times \Fg^i \times
    \Fg^j \times \Fg^k .
\label{4.18}
\ee

\nit
The action constructed from this prepotential becomes

\be
S_{int}\left[ T \right]\, =\, \int d\tau\, \lh \frac{ie}{2} T_{ijk}(x) \eta^i
   \eta^j f^k + \frac{1}{3!} \chi T_{ijk}(x) \eta^i \eta^j \eta^k +
   \frac{e}{3!} \ps^{\mu} \pl_{\mu} T_{ijk}(x) \eta^i \eta^j \eta^k \rh
\label{4.19}
\ee

\nit
Comparison with the action $S_{pot}[U]$ shows, that this action represents
the coupling to non-abelian scalar fields, where the generators of the group
are again expressed in terms of the rank-$r$ Grassmann algebra. Extensions of
these results to higher-order forms are straightforward.

\section{Applications}{\label{S.5}}

The actions constructed above can be used to describe spinning particles
in a $d$-dimensional target space-time in various kinds of background
fields: scalar fields, abelian and non-abelian vector fields,
anti-symmetric tensor fields, rank-3 anti-symmetric torsion, etc. (Note that
in four-dimensional space-time the rank-3 anti-symmetric tensor is dual
to an axial vector field.) In this section I discuss some special examples
which are particularly useful in physics applications.
\vs

\nit
{\em 1.\ Yukawa coupling.} One of the simpler cases is that of a spinning
particle in Minkowski space-time interacting with a scalar field.
This situation is described by the kinetic action with $g_{\mu\nu} =
\eta_{\mu\nu}$, the Minkowski metric, extended with the action $S_{ferm}$
for the internal fermion variables in a flat background $(K_{ij} = \del_{ij})$
and a potential term $\lb S_{pot}$, where $\lb$ is the coupling constant:

\be
S_{Yuk}\, =\, m S_{kin}\left[ \eta_{\mu\nu} \right]\, +\,
          S_{ferm}\left[ \del_{ij} \right]\, -\, \lb S_{pot}\left[ U \right].
\label{5.0}
\ee

\nit
The full component action is

\be
\ba{lll}
S_{Yuk} & = & \dsp{ \int d\tau \lh \frac{m}{2e} \dot{x}_{\mu}^2 +
  \frac{im}{2}\, \ps_{\mu} \dot{\ps}^{\mu} + \frac{im}{e}\, \chi \ps_{\mu}
  \dot{x}^{\mu} + \frac{i}{2}\, \eta_i \dot{\eta}^i +
  \frac{e}{2}\, f_i^2\rd } \\
  & & \\
  & & \dsp{ \ld \hspace{3em}
      -\, \lb e U_i(x) f^i\, +\, i \lb U_i(x) \chi \eta^i\, +\,
      i \lb e \ps^{\mu} \pl_{\mu} U_i(x) \eta^i \rh }
\ea
\label{5.1}
\ee

\nit
The auxiliary variables $f^i$ can be eliminated using their algebraic
Euler-Lagrange equation

\be
f_i\, =\, \lb U_i(x).
\label{5.2}
\ee

\nit
This gives the result

\be
\ba{lll}
S_{Yuk} & = & \dsp{ \int d\tau \lh \frac{m}{2e} \dot{x}_{\mu}^2 +
  \frac{im}{2}\, \ps_{\mu} \dot{\ps}^{\mu} + \frac{im}{e}\, \chi \ps_{\mu}
  \dot{x}^{\mu} + \frac{i}{2}\, \eta_i \dot{\eta}^i -
  \frac{e}{2}\, \lb^2 U_i^2 \rd } \\
  & & \\
  & & \dsp{ \ld \hspace{3em} +\, i \lb U_i(x) \chi \eta^i\,
            +\, i \lb e \ps^{\mu}\pl_{\mu} U_i(x) \eta^i \rh. }
\ea
\label{5.3}
\ee

\nit
The constraints from varying the action with respect to the gauge variables
are

\be
\ba{rcl}
\dsp{ m^2 \lh \dot{x}_{\mu} + i \chi \ps_{\mu} \rh^2 } & = & \dsp{
  - m e^2 \lh \lb^2 U_i^2 - 2 i \lb \ps^{\mu} \pl_{\mu} U_i \eta^i \rh, } \\
  & & \\
\dsp{ m \dot{x}_{\mu} \ps^{\mu} + e \lb U_i \eta^i } & = & 0.
\ea
\label{5.3.1}
\ee

\nit
The model describes a spinning particle in a relativistic scalar potential
$\lb^2 U_i^2(x)$, which may be dynamical. If this field has a vacuum
expectation value $\lb^2 \langle U_i^2 \rangle = m c^2$, it generates a mass
for the particle, showing it can act as a Higgs field. This mechanism of
generating mass dynamically is an alternative to adding the non-linear
multiplet action. However, in some sense the two mechanisms are the same,
because the action for the linear fermionic multiplet $(\eta, f)$ becomes
identical with the non-linear multiplet action $S_{nl}(\sg;c)$ if one imposes
the constraint that $f = c$, a constant.
\vs

\nit
{\em 2.\ Yang-Mills coupling.} A very interesting application from the point
of view of particle physics is the case of a spinning particle (e.g., a quark
or lepton) coupled to a vector gauge field $V_{\mu}(x)$ \ct{BCL,jarvis,jw14},
a supersymmetric generalization of Wong's model \ct{wong}. Again, I consider
ordinary Minkowski space-time and a flat internal space-time. Then adding the
vector action:

\be
S_{gauge}\, =\, m S_{kin}\left[ \eta_{\mu\nu} \right] +
                S_{ferm}\left[ \del_{ij} \right] - g S_{YM}\left[ V \right],
\label{5.4}
\ee

\nit
and eliminating the auxiliary variable $f^i$ by its Euler-Lagrange equation

\be
 f_{i}\, =\, ig V_{ij\mu}(x) \eta^j \ps^{\mu},
\label{5.5}
\ee

\nit
the component action reads

\be
S_{gauge}\, =\, \int d\tau \lh \frac{m}{2e} \dot{x}_{\mu}^2 +
  \frac{im}{2}\, \ps_{\mu} \dot{\ps}^{\mu} + \frac{im}{e}\, \chi \ps_{\mu}
  \dot{x}^{\mu} + \frac{i}{2}\, \eta_i \dot{\eta}^i + g \bar{V}_{\mu}
  \dot{x}^{\mu} - \frac{ige}{2}\, \bar{F}(V)_{\mu\nu} \ps^{\mu} \ps^{\nu} \rh.
\label{5.6}
\ee

\nit
For convenience I have introduced here the Grassmann-algebra valued gauge
field

\be
\bar{V}_{\mu}\, =\, - \frac{i}{2}\, \eta^i \eta^j V_{ij\mu},
\label{5.7}
\ee

\nit
and similarly for the field-strength:

\be
\bar{F}(V)_{\mu\nu}\, =\, - \frac{i}{2}\, \eta^i \eta^j F_{ij\mu\nu}(V)\, =\,
  \pl_{\mu} \bar{V}_{\nu} - \pl_{\nu} \bar{V}_{\mu}
  - g\, \overline{\left[ V_{\mu}, V_{\nu} \right]}.
\label{5.8}
\ee

\nit
The equation of motion for a particle in a non-abelian background
gauge field then becomes

\be
m \frac{d^2 x^{\mu}}{d\ttau^2}\, =\,
  g \bar{F}(V)_{\:\:\nu}^{\mu} \frac{dx^{\nu}}{d\ttau}\,
  - \frac{ig}{2}\, D^{\mu} \bar{F}(V)_{\lb\nu} \ps^{\lb} \ps^{\nu},
\label{5.9}
\ee

\nit
where $d\ttau = e d\tau$ and $D^{\mu}$ is the gauge-covariant derivative.
The first term represents the non-abelian Lorentz force, the second one the
Stern-Gerlach term responsible for non-abelian spin--orbit interactions
\ct{jw1a}.
\vs

\nit
{\em 3.\ Gravity.}  The actions above can be easily generalized to include
gravity, by using a general curved-space metric $g_{\mu\nu}(x)$ in the kinetic
multiplet rather than the Minkowski metric $\eta_{\mu\nu}$. The internal-space
metric $K_{ij}(x)$ however remains flat. The only new feature is then to
change the kinetic terms to the general form $S_{kin}[g]$, eq.(\ref{4.5}).
\vs

\nit
{\em 4.\ Anti-symmetric tensor coupling.} As a final example of the coupling
of spinning particles to external fields we consider anti-symmetric
rank-2 tensor fields $B_{i\mu\nu}(x)$ in curved space-time as well as internal
space. The action to use is

\be
S_{tensor}\, =\, m S_{kin}[g] + S_{ferm}[K] - y S_{even}[B],
\label{5.10}
\ee

\nit
where $y$ is a coupling constant. The auxiliary variables $f^i$ now satisfy
the equation

\be
K_{ij} f^j\, =\, \frac{i}{2}\, \lh \ps^{\mu} \pl_{\mu} K_{ij} \eta^j
   - y B_{i\mu\nu} \ps^{\mu} \ps^{\nu} \rh.
\label{5.11}
\ee

\nit
To solve it, we assume that $K_{ij}(x)$ is invertible. Elimination of
the auxiliary variables from the action then gives the component result

\be
\ba{lll}
S_{tensor} & = & \dsp{ \int d\tau\, \lh \frac{m}{2e}\, g_{\mu\nu}\,
  \dot{x}^{\mu} \dot{x}^{\nu} + \frac{im}{2}\, g_{\mu\nu}\, \psi^{\mu}
  D\psi^{\nu} + \frac{im}{e}\, g_{\mu\nu}\, \chi \psi^{\mu}
  \dot{x}^{\nu} + \frac{i}{2}\, K_{ij}\, \eta^i \dot{\eta}^j \rd }\\
  & & \\
  & & \dsp{ -\, i \eta^i\, B_{i\mu\nu}\, \ps^{\mu} \dot{x}^{\nu} +
  \frac{1}{2}\, \chi \eta^i\, B_{i\mu\nu}\, \ps^{\mu} \ps^{\nu} +
  \frac{e}{2}\, \eta^i\, F_{i \lb\mu\nu}(B)\, \ps^{\lb}\ps^{\mu}\ps^{\nu} } \\
  & & \\
  & & \dsp{ +\, \frac{e}{8}\, \eta^i \eta^j\, \lh \pl_{\mu} K \cdot K^{-1}
  \cdot \pl_{\nu} K \rh_{ij}\, \ps^{\mu} \ps^{\nu} - \frac{ye}{4}\,
  \eta^i \lh \pl_{\lb} K \cdot  K^{-1} \rh_i^{\:\:j} B_{j\mu\nu}\,
  \ps^{\mu} \ps^{\nu} \ps^{\lb}  } \\
  & & \\
  & & \dsp{ \ld -\, \frac{y^2 e}{8}\, B_{\mu\nu} \cdot K^{-1} \cdot
  B_{\kg\lb}\, \ps^{\mu} \ps^{\nu} \ps^{\kg} \ps^{\lb} \rh . }
\ea
\label{5.12}
\ee

\nit
Here $F_{i\mu\nu\lb}(B) = 1/3\, (\pl_{\lb} B_{i\mu\nu} + \pl_{\nu} B_{i\lb\mu}
+ \pl_{\mu} B_{i\nu\lb} )$ is the field-strength of the anti-symmetric tensor
field. When the internal metric is flat: $K_{ij}(x) = \del_{ij}$, considerable
simplifications occur and the whole third line vanishes. In four dimensions
the product $\ps^{\mu}\ps^{\nu}\ps^{\kg}\ps^{\lb}$ is proportional to
$\ve^{\mu\nu\kg\lb}$, and the last term is of the form $B \cdot K^{-1} \cdot
\tilde{B}$, where the tilde denotes the dual tensor.

\section{Space-time supersymmetry}{\label{S.6}}

In all previous examples the fermionic multiplets were used to represent
internal degrees of freedom, connected with rigid or local internal
symmetries. I conclude this paper with an application where the
extra fermionic variables represent space-time degrees of freedom.
This example is the spinning superparticle \ct{lukjw,luk}, which possesses
both (local) world-line and (rigid) target-space supersymmetry \ct{GS}.

For simplicity, I consider only space-times which allow Majorana spinors
($d = 2,3,4$ mod 8). In such a space-time one can define, in addition to the
usual co-ordinate multiplets $\Sg^{\mu}$, a spinor of real fermionic
supermultiplets

\be
\Ps_a\, =\, \lh \th_a, h_a \rh,
\label{6.1}
\ee

\nit
with $a = 1,...,2^{[\frac{d}{2}]}$. More generally, in  an arbitrary spinor
basis we do not require reality, but the Majorana condition

\be
\Ps\, =\, C \bar{\Ps},
\label{6.2}
\ee

\nit
where $\bar{\Ps} = \Ps^{\dagger} \gam_0$ is the Pauli conjugate spinor, and
$C$ is the charge conjugation matrix. Then the components $(\th_a, h_a)$
define an anti-commuting and a commuting Majorana spinor in then target
space-time, respectively. The super-derivative of this spinor of multiplets
is defined as in eq.(\ref{3.4}).

Introducing the Dirac matrices $\gam^{\mu}$ in the $d$-dimensional target
space-time, I next construct a $d$-vector of composite spinor multiplets

\be
\Og^{\mu}\, =\, \cD \Sg^{\mu} - \cD \bar{\Ps} \gam^{\mu} \Ps.
\label{6.3}
\ee

\nit
The components of these spinor multiplet are

\be
\Og^{\mu}\, \equiv\, \lh \og^{\mu}, \Pi^{\mu} \rh\,
  =\, \lh \ps^{\mu} - \bar{h} \gam^{\mu} \th, \frac{1}{e}\, \cD_{\tau} x^{\mu}
+
  \frac{i}{e}\, \cD_{\tau} \bar{\th} \gam^{\mu} \th - \bar{h} \gam^{\mu} h \rh.
\label{6.4}
\ee

\nit
{}From the spinor supermultiplets $\Og^{\mu}$ it is straightforward to
construct a fermionic prepotential which is a Lorentz scalar in
target space-time:

\be
\Fg \left[ \Ps \right]\, =\,
           \frac{1}{2}\, \eta_{\mu\nu}\, \cD \Og^{\mu} \times \Og^{\nu}.
\label{6.5}
\ee

\nit
The component action derived from this prepotential is

\be
\ba{lll}
S_{super} & = & \dsp{ \int d\tau\, \left[ \frac{1}{2e}\, \lh \dot{x}_{\mu}
  - i \bar{\th} \gam_{\mu} \dot{\th} - e \bar{h} \gam_{\mu} h \rh ^2
  + \frac{i}{2}\, \lh \ps_{\mu} - \bar{h} \gam_{\mu} \th \rh \frac{d}{d\tau}
  \lh  \ps_{\mu} - \bar{h} \gam_{\mu} \th \rh  \rd }\\
 & & \\
 & & \dsp{ \ld \hspace{3em} +\, \frac{i}{e}\, \chi \lh  \ps_{\mu} -
  \bar{h} \gam_{\mu} \th \rh \lh \dot{x}_{\mu} -
  i \bar{\th} \gam_{\mu} \dot{\th} - e \bar{h} \gam_{\mu} h \rh \right]. } \\
 & & \\
 & = & \dsp{ \int d\tau\, \left[ \frac{e}{2}\, \Pi_{\mu}^2 + \frac{i}{2}\,
        \og_{\mu} \dot{\og}^{\mu} \right]. }
\ea
\label{6.6}
\ee

\nit
A superfield derivation of this action has been presented in \ct{kavalov}.
Similar models in two-dimensional space-time describing spinning
strings were constructed in \ct{Gates,Brooks,Fisch}. We observe that $h$
is an auxiliary commuting Majorana spinor. Contrary to our previous actions,
in $S_{super}$ these auxiliary variables in general have a cubic and a quartic
term, of the form $ \gam_{\mu} h \bar{h} \gam^{\mu} h$ and $(\bar{h} \gam_{\mu}
h)^2$. However, owing to the Fierz identities these terms vanish in
four-dimensional space-time, where the auxiliary variables only appear
quadratically.

The action $S_{super}$ has a huge number of symmetries. Except for local
worldline supersymmetry, I mention rigid target-space supersymmetry, under
which the gauge multiplet $(e, \chi)$ is inert, whilst the linear multiplets
$\Sg^{\mu}$ and $\Ps_a$ transform with an anti-commuting Majorana spinor
parameter $\eps$:

\be
\ba{ll}
\del x^{\mu} = - i \bar{\th} \gam^{\mu} \eps, & \del \ps^{\mu} = \bar{h}
                                                \gam^{\mu} \eps, \\
 & \\
\del \th = \eps, & \del h = 0.
\ea
\label{6.7}
\ee

\nit
These transformations imply that the components of the multiplet $\Og^{\mu} =
(\og^{\mu}, \Pi^{\mu})$ in eq.(\ref{6.4}) are invariant: $\del \og^{\mu} =
\del \Pi^{\mu} = 0$.

Then there is the Siegel invariance with anti-commuting spinor parameter $\kg$
on the worldline, which takes the form

\be
\ba{ll}
\del x^{\mu} = i \bar{\th} \gam^{\mu} \gam \cdot \Pi \kg, & \dsp{
\del \ps^{\mu} = \frac{2i}{e}\, \bar{h} \gam^{\mu} \th \dot{\bar{\th}} \kg
                 + \bar{h} \gam^{\mu} \gam \cdot \Pi \kg, } \\
 & \\
\del \th = \gam \cdot \Pi \kg, & \dsp{
\del h = - \frac{2i}{e}\, h \dot{\bar{\th}} \kg, }\\
 & \\
\del e = 4i \dot{\bar{\th}} \kg, & \del \chi = 0.
\ea
\label{6.8}
\ee

\nit
Under these variations the components $(\og^{\mu}, \Pi^{\mu})$ transform as

\be
\del \Pi^{\mu} = \frac{2i}{e}\, \dot{\bar{\th}} \gam^{\mu} \gam \cdot \Pi \kg
               - \frac{4i}{e}\, \dot{\bar{\th}} \kg\, \Pi^{\mu},   \hspace{3em}
\del \og^{\mu} = 0.
\label{6.9}
\ee

\nit
In addition there is a bosonic counterpart of the Siegel invariance \ct{lukjw}
with commuting spinor parameter $\ag$:

\be
\ba{ll}
\del x^{\mu} = 0, & \del \ps^{\mu} = \bar{\th} \gam^{\mu} \gam \cdot \Pi \ag\,
                    -\, 2 \bar{\th} \gam^{\mu} h \bar{h} \ag, \\
 & \\
\del \th = 0, & \del h = \gam \cdot \Pi \ag\, -\, 2 h \bar{h} \ag, \\
 & \\
 \del e = - 4 e \bar{h} \ag, & \del \chi = 0,
\ea
\label{6.10}
\ee

\nit
resulting in

\be
\del \Pi^{\mu} = 2 \bar{\ag} \gam^{\mu} \gam \cdot \Pi h, \hspace{3em}
\del \og^{\mu} = 0.
\label{6.11}
\ee

\nit
Still other symmetries can be found for the massless spinning superparticle,
which I do not discuss here. If one assigns the space-time supersymmetry
transformation $\del \sg = 0$ to the non-linear fermion multiplet, addition
of the mass term $S_{grav}$ respects local world-line supersymmetry and
space-time supersymmetry. However, in this case the Siegel transformations
and their bosonic extension are no longer invariances of the model.
\vs

%%%%%%%%%%%%%%%%%%%%%%%%%%%%%%%%%%%%%%%%%%%%%%%%%%%%%%%%%%%%%%%%%%%%%%%%%%%%%%%
\np

\nit
{\bf Acknowledgement}
\vs

\nit
The research described in this paper is supported in part by the
Human Capital and Mobility program of the European Union through the
network on Constrained Dynamical Systems.

\end{document}